\documentclass{article}

\usepackage[pdftex]{graphicx}
 
 \usepackage{spconf,amsmath,subcaption, makecell, array}
 \usepackage{nicematrix}
\usepackage{enumitem}
\usepackage{algorithmic}
\usepackage{algorithm}
\usepackage{graphicx}
\usepackage[2]{editing}
\newcolumntype{L}[1]{>{\raggedright\let\newline\\\arraybackslash\hspace{0pt}}m{#1}}
\newcolumntype{C}[1]{>{\centering\let\newline\\\arraybackslash\hspace{0pt}}m{#1}}
\newcolumntype{R}[1]{>{\raggedleft\let\newline\\\arraybackslash\hspace{0pt}}m{#1}}
\makeatletter
\newcommand{\ssymbol}[1]{^{\@fnsymbol{#1}}}
\makeatother

\ninept

\title{Study of Robust Adaptive Beamforming Algorithms Based on Power Method Processing and Spatial Spectrum Matching }
\name{Saeed Mohammadzadeh$^{\star}$ \qquad Vítor H. Nascimento$^{\star}$ \qquad Rodrigo C. de Lamare$^{\dagger}$ \qquad Osman Kukrer$^{\dag\dag}$\vspace{-0.6em}}

\address{$^{\star}$ PSI, University of Sao Paulo, Brazil, 
$^{\dagger}$ PUC-Rio, Rio de Janeiro, Brazil, $^{\dag\dag}$ EEE, EMU, Turkey \vspace{-0.05em}}
\begin{document}
%
\maketitle
\begin{abstract}
Robust adaptive beamforming (RAB) based on interference-plus-noise covariance (INC) matrix reconstruction can experience performance degradation when model mismatch errors exist, particularly when the input signal-to-noise ratio (SNR) is large. In this work, we devise an efficient RAB technique for dealing with covariance matrix reconstruction issues. The proposed method involves INC matrix reconstruction using an idea in which the power and the steering vector of the interferences are estimated based on the power method. Furthermore, spatial match processing is computed to reconstruct the desired signal-plus-noise covariance matrix. Then, the noise components are excluded to retain the desired signal (DS) covariance matrix. A key feature of the proposed technique is to avoid eigenvalue decomposition of the INC matrix to obtain the dominant power of the interference-plus-noise region. Moreover, the INC reconstruction is carried out according to the definition of the theoretical INC matrix. Simulation results are shown and discussed to verify the effectiveness of the proposed method against existing approaches. 
\end{abstract}
\section{Introduction}
\label{sec:intro}
Adaptive beamforming methods have been applied in wireless communications, sonar, and radar due to their interference mitigation capability  \cite{van2004detection}. However, under non-ideal conditions such as finite data samples and mismatches between the presumed and true steering vector (SV) the performance of adaptive beamformers degrades substantially.  
Several robust adaptive beamforming (RAB) techniques have been proposed to enhance robustness against the aforementioned mismatches, such as the linearly constrained minimum variance (LCMV) beamformer \cite{monzingo2004introduction}, diagonal loading (DL) \cite{kukrer2014generalised} and other regularization strategies \cite{l1stap,l1cg,rdstap,spstap,rmmseprec,siprec,srmmse},  the subspace-based beamformers  \cite{huang2012modified,jidf,jio,jiostap,cgbf,sjidf,wlmwf,wljio,jiodoa,rrdoa,mskaesprit}, the worst case-based technique \cite{vorobyov2003robust,rccm,rdcapon}, the probabilistically constrained approach in \cite{vorobyov2008relationship} and the modified robust Capon beamformer in \cite{mohammadzadeh2018modified}. 
Hence, the development of low-complexity RAB approaches has been a very active research topic in recent years. Nevertheless, a major cause of performance degradation in adaptive beamforming is the presence of the desired signal (DS) component in the training data, especially at high SNR.

To address this issue, many works tried to remove the signal-of-interest (SOI) components by reconstruction of the interference-plus-noise covariance (INC) matrix instead of using the sample covariance matrix (SCM). In \cite{gu2012robust}, the INC matrix is reconstructed by integrating the nominal SV and the corresponding Capon spectrum over the entire angular sector except the region near the SOI. Several categories of INC matrix-based beamformers were then proposed, such as the beamformer in \cite{chen2015robust}, which relies on a correlation coefficient method, the computationally efficient algorithms via low complexity reconstruction in \cite{ruan2014robust,ruan2016,lrcc}, the sparse based method in \cite{gu2014robust}, the subspace-based algorithm in \cite{mohammadzadeh2018adaptive}, an approach based on spatial power spectrum sampling (SPSS) \cite{zhang2016interference}, and  methods based on coprime arrays in \cite{7929337,sparray1}, covariance matrices \cite{covbf} and convolutional networks \cite{rbfconv}. The beamformer in \cite{yuan2017robust} proposes a method in which each interference SV is estimated by the vector lying within the intersection of two subspaces while the algorithm in \cite{chen2018adaptive} constructs an INC matrix from the signal-interference subspace. The beamformer in \cite{gu2019adaptive} exploits orthogonal subspaces to eliminate the component of the SOI from the angle-related bases. The method in \cite{zhu2020robust} uses the orthogonality of subspaces to reconstruct the INC matrix while in \cite{MEPSalgorithm} a robust beamformer is proposed based on the principle of maximum entropy power spectrum (MEPS) to reconstruct the INC and the DS covariance matrices. Recently, an adaptive beamforming based on the idea of reconstructing the autocorrelation sequence  of a random process from a set of measured data was reported in \cite{mohammadzadeh2021robust}.\\
\indent In this paper, we develop an effective RAB approach based on power method processing and spatial spectrum matching (PMP-SSM) to reconstruct the INC matrix, which aims to reconstruct more precisely the INC and DS matrices. The essence of the idea is that the power and the SV of the desired signal and of the interferences are estimated by the eigenvalues and eigenvectors lying respectively within the interval of the SOI and the interference angular regions. In order to accomplish this, we devise a simple approach based on the power method \cite{ford2014numerical} where a simple iteration strategy is utilized for computing the dominant eigenvalues and corresponding eigenvectors. A key aspect of the proposed technique is to avoid the eigenvalue decomposition (EVD) required to reconstruct the INC matrix. In addition, an effective processing based on matching spectrum is developed to reconstruct the DS covariance matrix and a new SV estimation of SOI is obtained over the SOI angular sector using the estimated covariance matrix with the noise components excluded. In the proposed SV estimation algorithm, little prior information such as the imprecise knowledge of the antenna array geometry and the angular sectors is required, and the knowledge of the presumed steering vector is not essential.

\section{Problem Background}
\label{sec:methodology}
Let us consider a uniform linear array (ULA) composed of $ M$ omnidirectional array elements. Assume that $L$ narrowband signals (one SOI and $ L-1$ interferences) impinge on the array from the directions $\lbrace \theta_i \rbrace_{l=1}^L$. The array received vector at time instant $ k$, denoted by $\mathbf{x}(k)= \mathbf{x}_\mathrm{s}(k)+\mathbf{x}_l(k)+\mathbf{x}_\mathrm{n}(k) $, can be modeled as  
\begin{align}\label{Received Data Vector}
\mathbf{x}(k)= s_1(k)\mathbf{a}_1+\sum_{l=2}^L s_l(k) \mathbf{a}_l+\mathbf{x}_\mathrm{n}(k),
\end{align}
where $\mathbf{x}_\mathrm{s}(k)$, $\mathbf{x}_l(k)$, and $\mathbf{x}_\mathrm{n}(k)$ $\in \mathcal{C}^{M \times 1}$ are statistically independent components representing the SOI, interferences, and sensor noise, respectively. $\mathbf{s}(k)=[s_1(k), \cdots, s_L(k)]^T \in  \mathcal{C}^{L \times 1} $ is the signal waveform vector where $(\cdot)^T$ denotes the transpose and $\mathbf{a}_l=\mathbf{a}(\theta_l) \in  \mathcal{C}^{M \times 1}$ is the SV associated with the $l$th source signal. $\mathbf{x}_\mathrm{n}(k)$ is assumed to be complex Gaussian noise vector with zero mean and covariance $\sigma^2_n \mathbf{I}_M$, and $\mathbf{I}_M$ stands for the $M\times M$ identity matrix.
Assuming that the SV $ \mathbf{a}(\theta_1) $ is known, then for a given beamformer weight vector $ \mathbf{w} $, the beamformer performance is measured by the output signal-to-interference-plus-noise ratio (SINR) as follows
\begin{align}\label{SINR}
\mathrm{SINR}=\sigma^{2}_1 |\mathbf{w}^\mathrm{H} \mathbf{a}(\theta_1)|^2 \big{/} \mathbf{w}^\mathrm{H} \mathbf{R}_\mathrm{i+n}\mathbf{w},
\end{align}
where $ \sigma^{2}_1 $ and $ \mathbf{R}_\mathrm{i+n}$ are the power of the DS and the INC matrix, respectively, and $ (\cdot)^\mathrm{H} $ stands for Hermitian transpose. Assuming that the interfering signals are independent, the covariance matrix of the received signal vector is given by
\begin{align}\label{Theoretical R}
    \mathbf{R}=  \sigma^{2}_1\mathbf{a}(\theta_1)\mathbf{a}^\mathrm{H}(\theta_1)+\sum_{l=2}^L \sigma^{2}_l\mathbf{a}(\theta_l)\mathbf{a}^\mathrm{H}(\theta_l)+\sigma^{2}_n\mathbf{I},   
\end{align}
where  $\sigma^2_n$ and $\sigma^2_l$ represent the power of white Gaussian noise and of each interference component, respectively. The problem of maximizing the SINR in (\ref{SINR}) can be cast as the following optimization
problem:
\begin{align}\label{MVDR}
\underset{{\mathbf{w}}}{\operatorname{min}}\ \mathbf{w}^\mathrm{H} \mathbf{R}_\mathrm{i+n} \ \mathbf{w}\ \hspace{.4cm} \mathbf{s.t.} \hspace{.4cm} \mathbf{w}^\mathrm{H} \mathbf{a}(\theta_1)=1.
\end{align} 
The solution to (\ref{MVDR}) yields the optimal beamformer given by
\begin{align}\label{optimal wegight vector}
\mathbf{w}_{\mathrm{opt}}=\mathbf{R}_\mathrm{i+n}^{-1} \mathbf{a}(\theta_1) \big{/} \mathbf{a}^\mathrm{H}(\theta_1) \mathbf{R}_\mathrm{i+n}^{-1}\mathbf{a}(\theta_1).
\end{align}
Moreover, the array covariance matrix $\mathbf{R}=\mathrm{E}\{\mathbf{x}(k)\mathbf{x}^\mathrm{H}(k)\}$ is also given by \cite{mohammadzadeh2019robust}
\begin{equation} \label{Theoretical R integral}
 \mathbf{R}=\mathbf{R}_\mathrm{i+n}+\mathbf{R}_\mathrm{s}= \int_{\Phi} P(\theta)\mathbf{a}(\theta)\mathbf{a}^\mathrm{H}(\theta) d\theta,  
\end{equation}
where $P(\theta)$ is the angular power spectrum of the signals, $\Phi=[\bar{\Theta} \cup \Theta_s]$ covers the union of the angular sectors of the DS and the interference-plus-noise signal,  $\bar{\Theta}=\bar{\Theta}_l \cup \bar{\Theta}_n$ ($\bar{\Theta}_l$ is the interference angular sector and $\bar{\Theta}_n$  denotes the region excluding the signal and interference) and of the DS region, $\Theta_s$ (obtained through low-resolution direction finding methods \cite{van2004detection}), while   $\mathbf{R}_\mathrm{s}=\sigma_1^2\mathbf{a}_1\mathbf{a}_1^\mathrm{H}$ is the theoretical DS covariance matrix. Since the exact INC matrix, $\mathbf{R}_\mathbf{i+n} $ is unavailable, it is replaced by the SCM, $\hat{\mathbf{R}}=(1/K)\sum_{t=1}^{K} \mathbf{x}(k)\mathbf{x}^\mathrm{H}(k) $, where  $K$ is the number of snapshots.
\section{Proposed INCPMP-SSM Algorithm}
\subsection{The INC matrix reconstruction}
We estimate the corresponding parameters of the INC matrix according to its definition as in \eqref{Theoretical R}, namely the SVs, powers of all interferences and noise variance. Using low-resolution direction finding methods \cite{somasundaram2014degradation} to estimate the directions-of-arrival (DoA) of all interferences would lead to certain look direction estimation errors. That is to say, the DoAs of all interferences always lie in some angular sectors and $\bar{\Theta}_l$, $l=2,3,\cdots,L$ is assumed as the angular sector in which the $l$th interference is located.\\ 
The essence of the proposed method to reconstruct the INC matrix is according to the approach in \cite{MEPSalgorithm}. Thus, we develop an idea which is based on the use of the maximum entropy power spectrum distribution over all possible directions and coarse estimates of the angular regions where the interference and noise lie: 
\begin{align}\label{Power of MEM}
\hat{P}(\theta)=\dfrac{1}{\epsilon_p  \vert  \mathbf{a}^\mathrm{H}(\theta)\hat{\mathbf{R}}^{-1} \mathbf{b}_1 \vert ^2},
\end{align}
where $\mathbf{b}_1=[\begin{smallmatrix}1 & 0 & \cdots & 0\end{smallmatrix}]^\mathrm{T}$,   $\epsilon_p=1/\mathbf{b}_1^\mathrm{T}\hat{\mathbf{R}}^{-1}\mathbf{b}_1$.
Utilizing \eqref{Theoretical R integral}, restricted to the angular sector $\bar{\Theta}$ and using the maximum entropy power estimate (\ref{Power of MEM}), the INC matrix can be reconstructed by numerically evaluating
\begin{align}\label{proposed Ri+n}
\hat{\mathbf{R}}_{\mathrm{i+n}}=\int_{\bar{\Theta}}\hat{P}(\theta)\mathbf{a}(\theta)\mathbf{a}^\mathrm{H}(\theta)d\theta.
\end{align}
Sampling $\bar{\Theta}$ uniformly with $Q \gg M$ sampling points 
spaced by $\Delta\theta$, (\ref{proposed Ri+n}) can be approximated by
\begin{align} \label{Summation}
\hat{\mathbf{R}}_{\mathrm{i+n}} \approx \sum_{i=1}^{Q} \dfrac{\mathbf{a}(\theta_i)\mathbf{a}^\mathrm{H}(\theta_i)}
{\epsilon_p  \vert  \mathbf{a}^\mathrm{H}(\theta_i)\hat{\mathbf{R}}^{-1} \mathbf{b}_1 \vert ^2} \Delta\theta,
\end{align}
which needs more computations (multiplications and summations). However, in the proposed method, instead of summation over the whole angular sector $\bar{\Theta}$ (requiring $Q$ sampling points) away from the SOI, we only make summation over the small angular sector $\bar{\Theta}_l$ (thus requiring less sampling points) of the $l$th interference to obtain the $l$th interference covariance matrix as
\begin{align} \label{ less Summation}
\hat{\mathbf{C}}_l \approx \sum_{lj=1}^{J_l} \dfrac{\mathbf{a}(\theta_{lj})\mathbf{a}^\mathrm{H}(\theta_{lj})}
{\epsilon_p  \vert  \mathbf{a}^\mathrm{H}(\theta_{lj})\hat{\mathbf{R}}^{-1} \mathbf{b}_1 \vert ^2} \Delta\theta,
\end{align}
where $\mathbf{a}(\theta_{lj})$ is the SV associated with $\lbrace \theta_{lj}\in \bar{\Theta}_{ld} \rbrace_{j=1}^{J_l}$, $\bar{\Theta}_{ld}$ is a discretization of the angular sector $\bar{\Theta}_l$ with $J_l\ll Q$ sampling points. This leads to tiny angular interval. Since only one interference signal is assumed in the uncertainty region ($\bar{\Theta}_l$), the principal eigenvector of the constructed matrix will be the SV we are looking for. To avoid the EVD processing, and reduce the complexity, the power method \cite{ford2014numerical} is employed to estimate the principal eigenvalue and the corresponding eigenvector of the reconstructed INC, $\hat{\mathbf{C}}_l$, based on the theorem below: \\
\indent \textbf{Theorem 1:} A Hermitian matrix $\mathbf{A} \in \mathcal{C}^{N \times N}$ has $N$ orthogonal eigenvectors $\mathbf{u}_1,\cdots, \mathbf{u}_N \ (\parallel \mathbf{u}_i \parallel_2=1$ for $i \in [1,\cdots,N]$). Assume its eigenvalues satisfy the relation $|\lambda_1| > |\lambda_2| \ge \cdots \ge |\lambda_N|$, and let $\mathbf{v}_0=\sum _{i=1}^N \alpha_i \mathbf{u}_i$ ($\alpha_i \ne 0$). Take the vector $\mathbf{v}_0$ as the initial vector, and form a vector sequence according to the power of $\mathbf{A}$ as
\begin{equation} \label{Power Method}
\mathbf{d}_k=
    \begin{cases}
    \mathbf{v}_k=\mathbf{A} \mathbf{v}_{k-1} \\ 
     m_k=\|\mathbf{v}_k\|_{\infty}\\
     \bar{\mathbf{v}}_k=\mathbf{v}_k/m_k, \ \ (k=1,2,\cdots)
    \end{cases}       
\end{equation}
then it holds that \cite{liu2015p}
\begin{align}
 \underset{{k \longrightarrow +\infty }}{\operatorname{lim}}\  \bar{\mathbf{v}}_k=\frac{\mathbf{u}_1}{\|\mathbf{u}_1\|_\infty}.
\end{align}
According to this power method theorem, the principal eigenvalue and corresponding eigenvector of matrix $\mathbf{A}$ can be computed.
Utilizing this theorem, we intend to find the largest eigenvalue and the principal eigenvector of the reconstructed interference matrices that retains as much as possible the power and SV of the interferences. By applying theorem 1 to the reconstructed matrix $\hat{\mathbf{C}}_l$ for every interference, the power of the interference and the steering vector are obtained. Therefore we obtain the $l$th interference covariance matrix term $\hat{\sigma}^2_l \hat{\mathbf{a}}_l \hat{\mathbf{a}}_l^\mathrm{H}$, the SVs $\hat{\mathbf{a}}_l$ and their powers $\hat{\sigma}^2_l$, ($l=2, 3, \cdots, L$). Since each region $\bar{\Theta}_l$ is small, $\hat{\mathbf{C}}_l$ will be approximately a rank-one matrix, and thus the convergence in Theorem 1 should be fast. Finally we obtain the interference covariance matrix $\sum _{l=2}^L \hat{\sigma}^2_l \hat{\mathbf{a}}_l \hat{\mathbf{a}}_l^\mathrm{H}$. 
In the noise region $\bar{\Theta}_n$, which is the complement of $\bar{\Theta}_l \cup \bar{\Theta}_s$, we employ the maximum entropy power spectrum distribution in \eqref{Power of MEM} to calculate the noise power as the average 
\begin{equation} \label{noise power}
    \bar{\sigma}_n^2=\dfrac{1}{T}\sum_{t=1}^T \dfrac{1}{\epsilon_p  \vert  \mathbf{a}^\mathrm{H}(\theta_t)\hat{\mathbf{R}}^{-1} \mathbf{b}_1 \vert ^2},
\end{equation}
where $\theta_t$ is a discrete sample point in $\bar{\Theta}_n$ , $T$ is the number of sample points. This approximation can be justified if we assume that $\mathbf{x}(k)$ is comprised of only complex Gaussian white noise, that is to say, the data covariance matrix becomes  $\hat{\mathbf{R}}=\sigma^2_n\mathbf{I}$. In this case,  based on \eqref{Power of MEM} the noise power can be obtained as follows: 
\begin{equation} \label{P hat}
    \hat{P}(\theta)=\dfrac{1}{\epsilon_p  \big \vert  \mathbf{a}^\mathrm{H}(\theta) \big(\sigma^2_n\mathbf{I} \big)^{-1} \mathbf{b}_1 \big \vert ^2}=\dfrac{(\sigma^2_n)^2}{\epsilon_p},
\end{equation}
where  $\epsilon_p=\frac{1}{\mathbf{b}_1^\mathrm{T}\hat{\mathbf{R}}^{-1}\mathbf{b}_1}=\sigma^2_n$. 
By replacing $\epsilon_p$ into \eqref{P hat} we can write
\begin{equation} \label{P hat final}
    \hat{P}(\theta)=\sigma^2_n.
\end{equation}
 \eqref{P hat final} implies that in the presence of only noise components, the power is composed of residual noise components which is same as the actual noise.
Then, from \eqref{noise power} and \eqref{P hat final} we can conclude that the actual noise power is estimated as  $\hat{\sigma}^2_n \approx \bar{\sigma}_n^2$.
\vspace{-0.05em}
\subsection{Desired signal steering vector estimation}
Assume that the DS lies in the angular sector $\Theta_s$ which is assumed to be distinguishable from the location of the interference signal. In the proposed INCPMP-SSM method, the maximum entropy power spectrum calculated by the reconstructed signal plus-noise covariance matrix, $\hat{\mathbf{R}}_\text{s+n}$ is denoted as $\hat{\mathbf{P}}_\text{s+n}(\theta)$ and that computed by the SCM, $\hat{\mathbf{R}}$ is depicted by $\mathbf{P}(\theta)$ are required to be matched in the angular sector $\Theta_s$. Besides, the spectrum corresponding to $\hat{\mathbf{R}}_\text{s+n}$ should approximate to the average noise power $\hat{\sigma}^2_n$ in the complement angular sector of $\Theta_s$, which is denoted as $\bar{\Theta}$.\\ 
The spectrum matching processing for signal-plus noise covariance matrix reconstruction has two main objectives: (i) minimize the difference between $\hat{\mathbf{P}}_\text{s+n}(\theta)$ and $\mathbf{P}(\theta)$ in the angular sector of $\Theta_s$. (ii) constrain the difference between the average noise power $\hat{\sigma}^2_n$ and the spatial spectrum of $\hat{\mathbf{P}}_\text{s+n}(\theta)$ in the angular sector $\bar{\Theta}$.\\
Since a covariance matrix is always positive semidefinite,  the positive semidefinite requirement of the reconstructed signal-plus-noise covariance matrix is guaranteed. The proposed INCPMP-SSM algorithm to obtain $\hat{\mathbf{R}}_\text{s+n}$ can be formulated as the optimization problem:
\begin{align}
  & \underset{{\hat{\mathbf{R}}_\text{s+n}}}{\operatorname{min}} \lVert \hat{\mathbf{P}}_\text{s+n}(\theta)- \mathbf{P}(\theta) \rVert_2 \nonumber \\ & \quad \text{s.t.} \lVert \hat{\mathbf{P}}_\text{s+n}(\theta)-\hat{\sigma}^2_n \rVert_2 < \zeta \nonumber \\ 
   & \qquad \quad \hat{\mathbf{R}}_\text{s+n}  \in \mathbf{S}^M_{+}
\end{align}
where $\zeta$ is a relatively small value, and $\mathbf{S}^M_+$ is a set of $M \times M$ positive semidefinite matrices. The above expression is rewritten as 
\vspace{-0.1em}
 \begin{align} \label{mini Rs}
      \underset{{\hat{\mathbf{R}}_\text{s+n}}}{\operatorname{min}} \Big(& \int_{\Theta_s}  \Big \vert \dfrac{1}{\epsilon_p \big \vert \mathbf{a}^\mathrm{H}(\theta)\hat{\mathbf{R}}_\text{s+n}^{-1} \mathbf{b}_1 \big \vert^2} - \dfrac{1}{\mathbf{a}^\mathrm{H}(\theta)\hat{\mathbf{R}}^{-1} \mathbf{a}(\theta)} \Big \vert^2 d\theta \Big)^{1/2} \nonumber \\
     &\text{s.t.} \int_{\bar{\Theta}} \Big ( \Big \vert \dfrac{1}{\epsilon_p \big \vert \mathbf{a}^\mathrm{H}(\theta)\hat{\mathbf{R}}_\text{s+n}^{-1} \mathbf{b}_1 \big \vert^2} -\hat{\sigma}^2_n \Big \vert^2 d\theta \Big )^{1/2} < \zeta \nonumber \\
     & \qquad \qquad \qquad \hat{\mathbf{R}}_\text{s+n}  \in \mathbf{S}^M_{+}
 \end{align}
To simplify the calculation of \eqref{mini Rs}, we choose a finite number of angles $\nu_j \in \Theta_s (j=1,2,\cdots,G)$ and $\phi_i \in \bar{\Theta} (i=1,2,\cdots,Q)$ to discretize the angular sector $\Theta_s$ and the complement angular sector $\bar{\Theta}$, respectively. Hence, \eqref{mini Rs} becomes
\begin{align} \label{Optimization}
    &\underset{{\hat{\mathbf{R}}_\text{s+n}}}{\operatorname{min}} \Bigg \lVert
    \begin{pmatrix}
    \dfrac{1}{\epsilon_p \big \vert \mathbf{a}^\mathrm{H}(\nu_1)\hat{\mathbf{R}}_\text{s+n}^{-1} \mathbf{b}_1 \big \vert^2} \\
    \vdots  \\
    \dfrac{1}{\epsilon_p \big \vert \mathbf{a}^\mathrm{H}(\nu_G)\hat{\mathbf{R}}_\text{s+n}^{-1} \mathbf{b}_1 \big\vert^2}  
    \end{pmatrix}
    -\begin{pmatrix}
    \dfrac{1}{\mathbf{a}^\mathrm{H}(\nu_1)\hat{\mathbf{R}}^{-1} \mathbf{a}(\nu_1)} \\
    \vdots  \\
    \dfrac{1}{\mathbf{a}^\mathrm{H}(\nu_G)\hat{\mathbf{R}}^{-1} \mathbf{a}(\nu_G)}  
    \end{pmatrix} \Bigg \rVert_2 \nonumber \\
    & \qquad \qquad \text{s.t.} \Bigg \lVert \begin{pmatrix}
    \dfrac{1}{\epsilon_p \big \vert \mathbf{a}^\mathrm{H}(\phi_1)\hat{\mathbf{R}}_\text{s+n}^{-1} \mathbf{b}_1 \big \vert^2} \\
    \vdots  \\
    \dfrac{1}{\epsilon_p \big \vert \mathbf{a}^\mathrm{H}(\phi_Q)\hat{\mathbf{R}}_\text{s+n}^{-1} \mathbf{b}_1 \big\vert^2}    
    \end{pmatrix} - \textbf{1}.\hat{\sigma}^2_n \Bigg \rVert_2 < \zeta \nonumber \\  & \qquad \qquad \qquad \qquad \qquad \hat{\mathbf{R}}_\text{s+n}  \in \mathbf{S}^M_{+}
\end{align}
where $\textbf{1}$ represents the $Q \times 1$ vector with all elements equal to one. However, \eqref{Optimization} is a nonconvex optimization problem since the objective function and the inequality constraint function are nonconvex functions. To convert \eqref{Optimization} to a convex optimization problem, we first use the reciprocal of the spatial spectrum to replace the original spatial spectrum, and then define $\mathbf{V}_s=[\mathbf{a}(\nu_1), \mathbf{a}(\nu_2),\cdots,\mathbf{a}(\nu_S)]$, $\mathbf{Y}_s=[\mathbf{a}(\phi_1), \mathbf{a}(\phi_2),\cdots,\mathbf{a}(\phi_Q)]$, and  $\mathbf{D}_s=\hat{\mathbf{R}}_\text{s+n}^{-1}$. Consequently, a new optimization problem is obtained as follows:
\begin{align} \label{final minimization}
  \underset{{\mathbf{D}}_s}{\operatorname{min}}& \big \lVert \text{diag}(\mathbf{V}^\mathrm{H}_s \mathbf{D}_s \mathbf{V}_s)-\text{diag}(\mathbf{V}^\mathrm{H}_s \hat{\mathbf{R}}^{-1} \mathbf{V}_s) \big \rVert_2 \nonumber \\
  &\text{s.t.} \big \lVert \text{diag}(\mathbf{Y}^\mathrm{H}_s \mathbf{D}_s \mathbf{Y}_s)- \mathbf{1}.(1/\hat{\sigma}^2_n) \big \rVert_2 < \zeta \nonumber \\
  & \qquad \qquad\qquad  \mathbf{D}_s  \in \mathbf{S}^M_{+}
\end{align}
where $\text{diag}(\cdot)$ is an operator which returns the diagonal vector of a matrix, $\zeta$ is a relatively small value which guarantees that the spatial spectrum of $\text{diag}(\mathbf{V}^\mathrm{H}_s \mathbf{D}_s \mathbf{V}_s)$ is close to the average noise power $\hat{\sigma}^2_n$ in the angular sector. We can find that \eqref{final minimization} is a convex optimization problem, and thus can be efficiently solved by CVX \cite{grant2008cvx}. By solving this optimization problem, the reconstructed signal-plus-noise covariance matrix $\mathbf{D}_s=\hat{\mathbf{R}}_\text{s+n}^{-1}$ is achieved. 
However, the solution of \eqref{final minimization} is the reconstructed signal plus-noise covariance matrix but not the covariance matrix of DS. Therefore, we have to remove the noise components from $\mathbf{D}_s=\hat{\mathbf{R}}_\text{s+n}^{-1}$. Recall that $\hat{\sigma}^2_n$ is the average noise power and assuming that the sensor noise is spatially white Gaussian noise, then the noise covariance matrix can be estimated by $\hat{\mathbf{R}}_\mathrm{n}=\hat{\sigma}^2_n \mathbf{I}$. Hence, the covariance matrix  of DS is calculated by
\begin{align} \label{Rs}
  \hat{\mathbf{R}}_\mathrm{s}= \mathbf{D}_s^{-1}- \hat{\sigma}^2_n \mathbf{I}.
\end{align}
According to the summary of the power method, \eqref{Power Method}, the principal eigenvalue $\hat{\sigma}_1^2$ and the eigenvector $\hat{\mathbf{a}}(\theta_1)$ of $\hat{\mathbf{R}}_\mathrm{s}$ can be computed. Note that the proposed SV estimation algorithm requires little prior information, such as imprecise knowledge of the array geometry and angular sectors, while knowledge of the assumed SV is not essential.
\vspace{-1.1em}
\begin{algorithm}
	\caption{Proposed PMP-SSM Adaptive Beamforming }\label{sobelcode}
	1: \textbf{Input:}\:Array received data vector $\lbrace \mathbf{{x}}(k) \rbrace_{k=1}^K $,\\
	2: \text{Initialize:}\: $m_0=1$, $\mathbf{u}_0=[1,1,\cdots,1]^T$, $\delta$; \\
	3:  Compute   ${\hspace{1em}}\hat{\mathbf{{R}}}=(1/K)\sum_{k=1}^{K} \mathbf{{x}}(k)\mathbf{{x}}^H(k)$;\\
	4: \textbf{For} $ l=2:L$ \\
	5:\quad \quad \ Construct $\hat{\mathbf{C}}_l$ using \eqref{ less Summation},\\
	6: \qquad Apply power method theorem in \eqref{Power Method} to $\hat{\mathbf{C}}_l$, \\
	7: \qquad  Obtain $\hat{\sigma}^2_l$ and $\hat{\mathbf{a}}_l$, \\
	8: \textbf{End For} \\
	9: Compute estimated noise power using \eqref{noise power}; \\
	10: Construct INC matrix as $ \hat{\mathbf{R}}_\mathrm{i+n}=\sum _{l=2}^L \hat{\sigma}^2_l \hat{\mathbf{a}}_l \hat{\mathbf{a}}_l^\mathrm{H} + \hat{\sigma}^2_n \mathbf{I}$.\\
	11: Compute $\mathbf{D}_s=\hat{\mathbf{R}}_\text{s+n}^{-1}$ using \eqref{final minimization}; \\
	12: Estimate the desired signal covariance matrix,$\hat{\mathbf{R}}_\mathrm{s}$  utilizing \eqref{Rs}, \\
	123: Apply power method theorem in \eqref{Power Method} to $\hat{\mathbf{R}}_\mathrm{s}$,\\
	14: Obtain $\hat{\sigma}^2_1$ and $\hat{\mathbf{a}}_1$. \\
	15: Design proposed beamformer using \eqref{optimal wegight vector},\\
	16: \textbf{Output:}\: Proposed beamforming weight vector $ \mathbf{w}_{\text{prop}} $.
\end{algorithm}
\section{Simulations}
In this section, a ULA with $ M=10 $ omnidirectional sensors is used.
It is assumed that there is one DS from the presumed direction  $\bar{\theta}_1=10^\circ $ while the uncorrelated interference signals are impinging from  $30^\circ$ and $50^\circ$. The input interference to noise ratios (INRs) of the two interferers are both set to 30 dB. The noise is regarded as complex Gaussian temporally and spatially white process with zero mean and unit variance. The proposed (INCPMP-SSM) method is compared with the beamformer in \cite{yuan2017robust} (INC-SUB),  the reconstruction-estimation based beamformer in \cite{gu2012robust} (INC-EST), the beamformer in \cite{zhu2020robust} (INC-OS), the beamformer in \cite{zhang2016interference} (INC-SPSS), the beamformer in \cite{zheng2018covariance} (INC-SV), the beamformer in \cite{MEPSalgorithm} (INC-MEPS) and the beamformer in \cite{du2010fully} (INC-FADL). In the INC-SV and INC-EST beamformers the number of sampling points for interference-plus-noise region is fixed at 200. In the beamformer INC-SV, the upper bound of the norm of the SV mismatch is set to $ \sqrt{0.1} $. In the proposed INCPMP-SSM method, $k=4$ iterations are used to compute the dominant eigenvector while we perform 100 Monte-Carlo runs. The dominant eigenvectors is fixed 7 in INC-OS. The energy percentage $\rho$ set as 0.9 in INC-SUB. The angular sector of the DS is set to be $ {\Theta}_s=[{\bar{\theta}_1}-5^\circ,{\bar{\theta}_1}+5^\circ] $ where the interference angular sector is $ \bar{\Theta}=[-90^\circ,{\bar{\theta}_1}-5^\circ)\cup({\bar{\theta}_1}+5^\circ,90^\circ] $. \\
\indent In the  example, we evaluate INCPMP-SSM in the presence of look direction and model mismatches due to the sensor displacement errors. In this example, it is assumed that the DS and the interferers are uniformly distributed in $ [-5^\circ,5^\circ] $ while the difference between the actual and assumed SV is modeled as array geometry errors, assuming the sensor position is drawn uniformly from $[-0.05,0.05]$ wavelength and the DoA of the DS and the actual sensor position changes from run to run while remaining constant over samples.

\begin{figure}[t]
	\centering
	\includegraphics[height=2.22in]{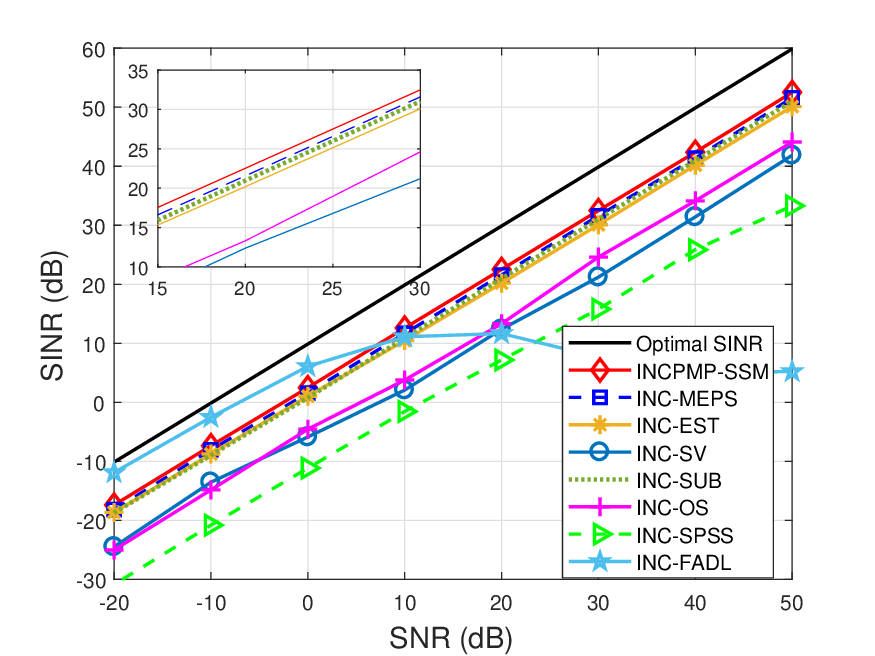}
\vspace{-0.95em}
	\caption{Output SINR versus input SNR}
	\label{SNR}
\end{figure}

In Fig.~\ref{SNR}, we compare the SINR performance versus the SNR where the number of snapshots is fixed at $K=50$. It is well-known that the theoretical INC is a linear combination of the SVs and powers of interferences, and the reconstructed INC in proposed method (INCPMP-SSM) is the linear combination of estimated powers $\hat{\sigma}^2_l$ and corresponding SVs $\hat{\mathbf{a}}_l$. It is demonstrated that the proposed INCPMP-SSM method outperforms the INC-MEPS and INC-SUB and it is better than INC-EST. From the results, it is observed that, because of random sensor position errors, there is an almost constant performance loss for INC-OS and INC-SPSS regardless of the input SNR. At SNRs higher than 0 dB, the INC-SV beamformer has a performance loss because of the look direction mismatch. On the other hand, the proposed INCPMP-SSM beamformer almost attains the optimal output SINR under these mismatches for all SNRs. It should be noted that, since the INC-FADL beamformer utilizes the assumed SV to compute the weight vector, its performance for low SNRs is better than the other beamformers.
\begin{figure}[t]
	\centering
	\includegraphics[height=2.22in]{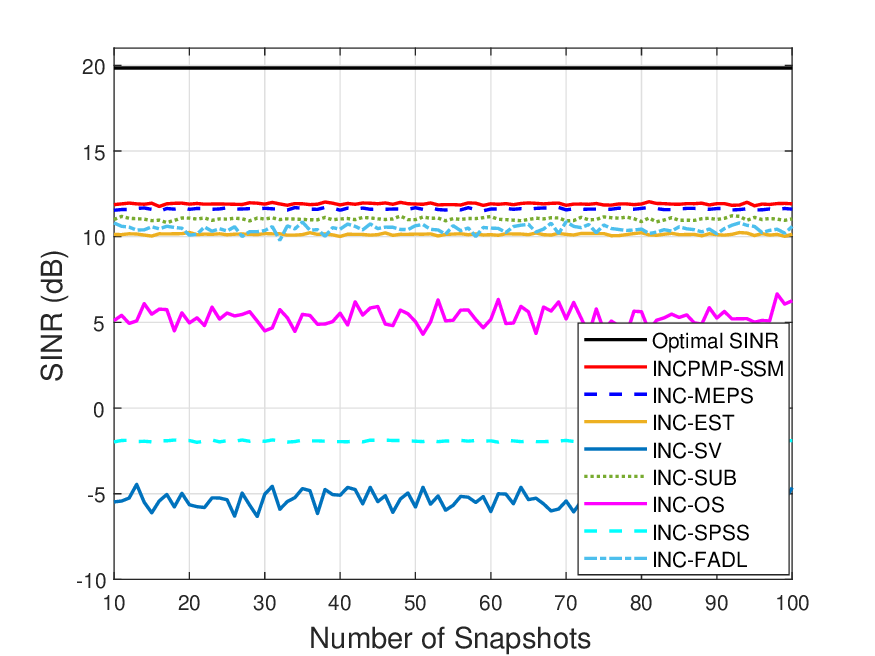}
\vspace{-0.95em}
	\caption{Output SINR versus number of snapshots}
	\label{Snapshots}
\end{figure}
\\ \indent In Fig.~\ref{Snapshots}, all tested beamformers are evaluated as the number of snapshots is increased at SNR=10 dB. The performance of the INCPMP-SSM method stems from its highly accurate estimate of the INC matrix with respect to the theoretical one, which enhances the robustness of the proposed INC against random look direction and array geometry errors over the snapshots. The results demonstrate that the number of snapshots does not significantly affect the output SINR of tested beamformers, and the INC-MEPS and INC-SUB  beamformers almost get the same performance as the INCPMP-SSM method. 

The proposed PMP-SSM beamformer is suitable for large sensor arrays \cite{mmimo,wence} and applications in communications \cite{spa,mfsic,mbdf,mbthp}, \cite{rmbthp,rsbd,rsthp,rapa,dynovs,comprec,zcprec} and distributed sensor systems \cite{damdc,rdistdiff,rmultitask} .

\section{Conclusion}
\label{sec:conclusion}
In this work, an efficient and accurate estimation of the INC matrix and DS steering vector has been proposed, where the eigenvalues and eigenvectors lying within the interval of the SOI and the interference angular regions are computed by the power method that estimates the actual power and SV of the interferences and desired signal. Furthermore, an efficient algorithm based on the matching spectrum is devised to reconstruct the DS covariance matrix and estimate the SV of the SOI. Simulation results have shown that the proposed INCPMP-SSM algorithm outperforms recently reported approaches.

\section{ACKNOWLEDGEMENTS}
This work was partially funded by FAPESP through the ELIOT project, grants 2018/12579-7 and 2019/19387-9.



\bibliographystyle{IEEEbib}
\bibliography{refs}

\begin{thebibliography}{10}

\bibitem{van2004detection}
Harry~L Van~Trees,
\newblock {\em Detection, Estimation, and Modulation Theory},
\newblock John Wiley \& Sons, New York, 2004.

\bibitem{monzingo2004introduction}
Robert~A Monzingo and Thomas~W Miller,
\newblock {\em Introduction to adaptive arrays},
\newblock Scitech publishing, 2004.

\bibitem{kukrer2014generalised}
O.~Kukrer and S.~Mohammadzadeh,
\newblock ``Generalised loading algorithm for adaptive beamforming in ulas,''
\newblock {\em IET Electronics Lett.}, vol. 50, no. 13, pp. 910--912, 2014.

\bibitem{l1stap}
Zhaocheng Yang, Rodrigo~C. de~Lamare, and Xiang Li,
\newblock ``$l_1$ -regularized stap algorithms with a generalized sidelobe canceler architecture for airborne radar,''
\newblock {\em IEEE Transactions on Signal Processing}, vol. 60, no. 2, pp. 674--686, 2012.

\bibitem{l1cg}
Z.~Yang, R.~C. de~Lamare, and X.~Li,
\newblock ``Sparsity-aware space–time adaptive processing algorithms with <i>l</i><sub>1</sub>-norm regularisation for airborne radar,''
\newblock {\em IET Signal Processing}, vol. 6, pp. 413--423(10), July 2012.

\bibitem{rdstap}
Xiaoye Wang, Zhaocheng Yang, Jianjun Huang, and Rodrigo~C. de~Lamare,
\newblock ``Robust two-stage reduced-dimension sparsity-aware stap for airborne radar with coprime arrays,''
\newblock {\em IEEE Transactions on Signal Processing}, vol. 68, pp. 81--96, 2020.

\bibitem{spstap}
Yang Zhaocheng, Rodrigo~C. de~Lamare, and Weijian Liu,
\newblock ``Sparsity-based stap using alternating direction method with gain/phase errors,''
\newblock {\em IEEE Transactions on Aerospace and Electronic Systems}, vol. 53, no. 6, pp. 2756--2768, 2017.

\bibitem{rmmseprec}
Victoria M.~T. Palhares, Andre~R. Flores, and Rodrigo~C. de~Lamare,
\newblock ``Robust mmse precoding and power allocation for cell-free massive mimo systems,''
\newblock {\em IEEE Transactions on Vehicular Technology}, vol. 70, no. 5, pp. 5115--5120, 2021.

\bibitem{siprec}
Yunlong Cai, Rodrigo C.~de Lamare, and Rui Fa,
\newblock ``Switched interleaving techniques with limited feedback for interference mitigation in ds-cdma systems,''
\newblock {\em IEEE Transactions on Communications}, vol. 59, no. 7, pp. 1946--1956, 2011.

\bibitem{srmmse}
Yunlong Cai, Rodrigo~C. de~Lamare, Lie-Liang Yang, and Minjian Zhao,
\newblock ``Robust mmse precoding based on switched relaying and side information for multiuser mimo relay systems,''
\newblock {\em IEEE Transactions on Vehicular Technology}, vol. 64, no. 12, pp. 5677--5687, 2015.

\bibitem{huang2012modified}
Fei Huang, Weixing Sheng, and Xiaofeng Ma,
\newblock ``Modified projection approach for robust adaptive array beamforming,''
\newblock {\em Signal Process.}, vol. 92, no. 7, pp. 1758--1763, 2012.

\bibitem{jidf}
Rodrigo~C. de~Lamare and Raimundo Sampaio-Neto,
\newblock ``Adaptive reduced-rank processing based on joint and iterative interpolation, decimation, and filtering,''
\newblock {\em IEEE Transactions on Signal Processing}, vol. 57, no. 7, pp. 2503--2514, 2009.

\bibitem{jio}
Rodrigo~C. de~Lamare and Raimundo Sampaio-Neto,
\newblock ``Reduced-rank adaptive filtering based on joint iterative optimization of adaptive filters,''
\newblock {\em IEEE Signal Processing Letters}, vol. 14, no. 12, pp. 980--983, 2007.

\bibitem{jiostap}
Rui Fa and Rodrigo~C. De~Lamare,
\newblock ``Reduced-rank stap algorithms using joint iterative optimization of filters,''
\newblock {\em IEEE Transactions on Aerospace and Electronic Systems}, vol. 47, no. 3, pp. 1668--1684, 2011.

\bibitem{cgbf}
L.~Wang and R.~C. de~Lamare,
\newblock ``Constrained adaptive filtering algorithms based on conjugate gradient techniques for beamforming,''
\newblock {\em IET Signal Processing}, vol. 4, pp. 686--697(11), December 2010.

\bibitem{sjidf}
Rui Fa, Rodrigo~C. de~Lamare, and Lei Wang,
\newblock ``Reduced-rank stap schemes for airborne radar based on switched joint interpolation, decimation and filtering algorithm,''
\newblock {\em IEEE Transactions on Signal Processing}, vol. 58, no. 8, pp. 4182--4194, 2010.

\bibitem{wlmwf}
Nuan Song, Rodrigo~C. de~Lamare, Martin Haardt, and Mike Wolf,
\newblock ``Adaptive widely linear reduced-rank interference suppression based on the multistage wiener filter,''
\newblock {\em IEEE Transactions on Signal Processing}, vol. 60, no. 8, pp. 4003--4016, 2012.

\bibitem{wljio}
Nuan Song, Waheed~Ullah Alokozai, Rodrigo~C. de~Lamare, and Martin Haardt,
\newblock ``Adaptive widely linear reduced-rank beamforming based on joint iterative optimization,''
\newblock {\em IEEE Signal Processing Letters}, vol. 21, no. 3, pp. 265--269, 2014.

\bibitem{jiodoa}
Lei Wang, Rodrigo~C. de~Lamare, and Martin Haardt,
\newblock ``Direction finding algorithms based on joint iterative subspace optimization,''
\newblock {\em IEEE Transactions on Aerospace and Electronic Systems}, vol. 50, no. 4, pp. 2541--2553, 2014.

\bibitem{rrdoa}
Linzheng Qiu, Yunlong Cai, Rodrigo~C. de~Lamare, and Minjian Zhao,
\newblock ``Reduced-rank doa estimation algorithms based on alternating low-rank decomposition,''
\newblock {\em IEEE Signal Processing Letters}, vol. 23, no. 5, pp. 565--569, 2016.

\bibitem{mskaesprit}
Silvio F.~B. Pinto and Rodrigo~C. de~Lamare,
\newblock ``Multistep knowledge-aided iterative esprit: Design and analysis,''
\newblock {\em IEEE Transactions on Aerospace and Electronic Systems}, vol. 54, no. 5, pp. 2189--2201, 2018.

\bibitem{vorobyov2003robust}
S.~A. Vorobyov, A.~B. Gershman, and Zhi-Quan Luo,
\newblock ``Robust adaptive beamforming using worst-case performance optimization: A solution to the signal mismatch problem,''
\newblock {\em IEEE Trans. on Signal Process.}, vol. 51, no. 2, pp. 313--324, 2003.

\bibitem{rccm}
Lukas Landau, Rodrigo~C. de~Lamare, and Martin Haardt,
\newblock ``Robust adaptive beamforming algorithms using the constrained constant modulus criterion,''
\newblock {\em IET Signal Processing}, vol. 8, no. 5, pp. 447--457, 2014.

\bibitem{rdcapon}
Samuel~D. Somasundaram, Nigel~H. Parsons, Peng Li, and Rodrigo~C. de~Lamare,
\newblock ``Reduced-dimension robust capon beamforming using krylov-subspace techniques,''
\newblock {\em IEEE Transactions on Aerospace and Electronic Systems}, vol. 51, no. 1, pp. 270--289, 2015.

\bibitem{vorobyov2008relationship}
Sergiy~A Vorobyov, Haihua Chen, and Alex~B Gershman,
\newblock ``On the relationship between robust minimum variance beamformers with probabilistic and worst-case distortionless response constraints,''
\newblock {\em IEEE Trans. on Signal Process.}, vol. 56, no. 11, pp. 5719--5724, 2008.

\bibitem{mohammadzadeh2018modified}
Saeed Mohammadzadeh and Osman Kukrer,
\newblock ``Modified robust {Capon} beamforming with approximate orthogonal projection onto the signal-plus-interference subspace,''
\newblock {\em Circuits, Systems, and Signal Process.}, pp. 1--18, 2018.

\bibitem{gu2012robust}
Y.~Gu and A.~Leshem,
\newblock ``Robust adaptive beamforming based on interference covariance matrix reconstruction and steering vector estimation,''
\newblock {\em IEEE Trans. on Signal Process.}, vol. 60, no. 7, pp. 3881--3885, 2012.

\bibitem{chen2015robust}
F.~Chen, F.~Shen, and J.~Song,
\newblock ``Robust adaptive beamforming using low-complexity correlation coefficient calculation algorithms,''
\newblock {\em IET Electronics Lett.}, vol. 51, no. 6, pp. 443--445, 2015.

\bibitem{ruan2014robust}
Hang Ruan and Rodrigo~C de~Lamare,
\newblock ``Robust adaptive beamforming using a low-complexity shrinkage-based mismatch estimation algorithm.,''
\newblock {\em IEEE Signal Process. Lett.}, vol. 21, no. 1, pp. 60--64, 2014.

\bibitem{ruan2016}
H.~{Ruan} and R.~C. {de Lamare},
\newblock ``Robust adaptive beamforming based on low-rank and cross-correlation techniques,''
\newblock {\em IEEE Trans. on Signal Process.}, vol. 64, no. 15, pp. 3919--3932, 2016.

\bibitem{lrcc}
Hang Ruan and Rodrigo~C. de~Lamare,
\newblock ``Distributed robust beamforming based on low-rank and cross-correlation techniques: Design and analysis,''
\newblock {\em IEEE Transactions on Signal Processing}, vol. 67, no. 24, pp. 6411--6423, 2019.

\bibitem{gu2014robust}
Yujie Gu, Nathan~A Goodman, Shaohua Hong, and Yu~Li,
\newblock ``Robust adaptive beamforming based on interference covariance matrix sparse reconstruction,''
\newblock {\em Signal Processing}, vol. 96, pp. 375--381, 2014.

\bibitem{mohammadzadeh2018adaptive}
Saeed Mohammadzadeh and Osman Kukrer,
\newblock ``Adaptive beamforming based on theoretical interference-plus-noise covariance and direction-of-arrival estimation,''
\newblock {\em IET Signal Process.}, vol. 12, no. 7, pp. 819--825, 2018.

\bibitem{zhang2016interference}
Zhenyu Zhang, Wei Liu, Wen Leng, Anguo Wang, and Heping Shi,
\newblock ``Interference-plus-noise covariance matrix reconstruction via spatial power spectrum sampling for robust adaptive beamforming,''
\newblock {\em IEEE Signal Process. Lett.}, vol. 23, no. 1, pp. 121--125, 2016.

\bibitem{7929337}
Chengwei Zhou, Yujie Gu, Shibo He, and Zhiguo Shi,
\newblock ``A robust and efficient algorithm for coprime array adaptive beamforming,''
\newblock {\em IEEE Transactions on Vehicular Technology}, vol. 67, no. 2, pp. 1099--1112, 2018.

\bibitem{sparray1}
Wanlu Shi, Yingsong Li, and Rodrigo~C. de~Lamare,
\newblock ``Novel sparse array design based on the maximum inter-element spacing criterion,''
\newblock {\em IEEE Signal Processing Letters}, vol. 29, pp. 1754--1758, 2022.

\bibitem{covbf}
Saeed Mohammadzadeh, Vítor~Heloiz Nascimento, Rodrigo C.~de Lamare, and Osman Kukrer,
\newblock ``Covariance matrix reconstruction based on power spectral estimation and uncertainty region for robust adaptive beamforming,''
\newblock {\em IEEE Transactions on Aerospace and Electronic Systems}, vol. 59, no. 4, pp. 3848--3858, 2023.

\bibitem{rbfconv}
Saeed Mohammadzadeh, Vítor~H. Nascimento, Rodrigo~C. de~Lamare, and Noushin Hajarolasvadi,
\newblock ``Robust beamforming based on complex-valued convolutional neural networks for sensor arrays,''
\newblock {\em IEEE Signal Processing Letters}, vol. 29, pp. 2108--2112, 2022.

\bibitem{yuan2017robust}
Xiaolei Yuan and Lu~Gan,
\newblock ``Robust adaptive beamforming via a novel subspace method for interference covariance matrix reconstruction,''
\newblock {\em Signal Processing}, vol. 130, pp. 233--242, 2017.

\bibitem{chen2018adaptive}
Peng Chen, Yixin Yang, Yong Wang, and Yuanliang Ma,
\newblock ``Adaptive beamforming with sensor position errors using covariance matrix construction based on subspace bases transition,''
\newblock {\em IEEE Signal Process. Lett.}, vol. 26, no. 1, pp. 19--23, 2018.

\bibitem{gu2019adaptive}
Yujie Gu and Yimin~D Zhang,
\newblock ``Adaptive beamforming based on interference covariance matrix estimation,''
\newblock in {\em 2019 53rd Asilomar Conference on Signals, Systems, and Computers}. IEEE, 2019, pp. 619--623.

\bibitem{zhu2020robust}
Xingyu Zhu, Xu~Xu, and Zhongfu Ye,
\newblock ``Robust adaptive beamforming via subspace for interference covariance matrix reconstruction,''
\newblock {\em Signal Processing}, vol. 167, pp. 107289, 2020.

\bibitem{MEPSalgorithm}
Saeed Mohammadzadeh, Vitor~H Nascimento, R.~C. de~Lamare, and Osman Kukrer,
\newblock ``Maximum entropy-based interference-plus-noise covariance matrix reconstruction for robust adaptive beamforming,''
\newblock {\em IEEE Signal Process. Lett.}, vol. 27, pp. 845--849, 2020.

\bibitem{mohammadzadeh2021robust}
Saeed Mohammadzadeh, Vitor~H Nascimento, Rodrigo~C De~Lamare, and Osman Kukrer,
\newblock ``Robust adaptive beamforming based on low-complexity discrete fourier transform spatial sampling,''
\newblock {\em IEEE Access}, 2021.

\bibitem{ford2014numerical}
William Ford,
\newblock {\em Numerical linear algebra with applications: Using MATLAB},
\newblock Academic Press, 2014.

\bibitem{mohammadzadeh2019robust}
Saeed Mohammadzadeh and Osman Kukrer,
\newblock ``Robust adaptive beamforming based on covariance matrix and new steering vector estimation,''
\newblock {\em Signal, Image and Video Processing}, vol. 13, no. 5, pp. 853--860, 2019.

\bibitem{somasundaram2014degradation}
Samuel~D Somasundaram and Andreas Jakobsson,
\newblock ``Degradation of covariance reconstruction-based robust adaptive beamformers,''
\newblock in {\em Sensor Signal Process. for Defence (SSPD), 2014}. IEEE, 2014, pp. 1--5.

\bibitem{liu2015p}
Fulai Liu, Ruiyan Du, Junping Guo, and Shouming Guo,
\newblock ``P-glrt algorithm for cooperative spectrum sensing,''
\newblock {\em Wireless Personal Communications}, vol. 81, no. 3, pp. 1079--1089, 2015.

\bibitem{grant2008cvx}
Michael Grant, Stephen Boyd, and Yinyu Ye,
\newblock ``Cvx: Matlab software for disciplined convex programming,'' 2008.

\bibitem{zheng2018covariance}
Zhi Zheng, Yan Zheng, Wen-Qin Wang, and Hongbo Zhang,
\newblock ``Covariance matrix reconstruction with interference steering vector and power estimation for robust adaptive beamforming,''
\newblock {\em IEEE Trans. on Vehicular Techn.}, vol. 67, no. 9, pp. 8495--8503, 2018.

\bibitem{du2010fully}
L.~Du, J.~Li, and P.~Stoica,
\newblock ``Fully automatic computation of diagonal loading levels for robust adaptive beamforming,''
\newblock {\em IEEE Trans. on Aerospace and Electronic Systems}, vol. 46, no. 1, pp. 449--458, 2010.

\bibitem{mmimo}
Rodrigo~C. de~Lamare,
\newblock ``Massive mimo systems: Signal processing challenges and future trends,''
\newblock {\em URSI Radio Science Bulletin}, vol. 2013, no. 347, pp. 8--20, 2013.

\bibitem{wence}
Wence Zhang, Hong Ren, Cunhua Pan, Ming Chen, Rodrigo~C. de~Lamare, Bo~Du, and Jianxin Dai,
\newblock ``Large-scale antenna systems with ul/dl hardware mismatch: Achievable rates analysis and calibration,''
\newblock {\em IEEE Transactions on Communications}, vol. 63, no. 4, pp. 1216--1229, 2015.

\bibitem{spa}
Rodrigo~C. De~Lamare and Raimundo Sampaio-Neto,
\newblock ``Minimum mean-squared error iterative successive parallel arbitrated decision feedback detectors for ds-cdma systems,''
\newblock {\em IEEE Transactions on Communications}, vol. 56, no. 5, pp. 778--789, 2008.

\bibitem{mfsic}
Peng Li, Rodrigo~C. de~Lamare, and Rui Fa,
\newblock ``Multiple feedback successive interference cancellation detection for multiuser mimo systems,''
\newblock {\em IEEE Transactions on Wireless Communications}, vol. 10, no. 8, pp. 2434--2439, 2011.

\bibitem{mbdf}
Rodrigo~C. de~Lamare,
\newblock ``Adaptive and iterative multi-branch mmse decision feedback detection algorithms for multi-antenna systems,''
\newblock {\em IEEE Transactions on Wireless Communications}, vol. 12, no. 10, pp. 5294--5308, 2013.

\bibitem{mbthp}
Keke Zu, Rodrigo~C. de~Lamare, and Martin Haardt,
\newblock ``Multi-branch tomlinson-harashima precoding design for mu-mimo systems: Theory and algorithms,''
\newblock {\em IEEE Transactions on Communications}, vol. 62, no. 3, pp. 939--951, 2014.

\bibitem{rmbthp}
Lei Zhang, Yunlong Cai, Rodrigo~C. de~Lamare, and Minjian Zhao,
\newblock ``Robust multibranch tomlinson–harashima precoding design in amplify-and-forward mimo relay systems,''
\newblock {\em IEEE Transactions on Communications}, vol. 62, no. 10, pp. 3476--3490, 2014.

\bibitem{rsbd}
Andre~R. Flores, Rodrigo~C. de~Lamare, and Bruno Clerckx,
\newblock ``Linear precoding and stream combining for rate splitting in multiuser mimo systems,''
\newblock {\em IEEE Communications Letters}, vol. 24, no. 4, pp. 890--894, 2020.

\bibitem{rsthp}
Andre~R. Flores, Rodrigo~C. De~Lamare, and Bruno Clerckx,
\newblock ``Tomlinson-harashima precoded rate-splitting with stream combiners for mu-mimo systems,''
\newblock {\em IEEE Transactions on Communications}, vol. 69, no. 6, pp. 3833--3845, 2021.

\bibitem{rapa}
André~R. Flores and Rodrigo~C. de~Lamare,
\newblock ``Robust and adaptive power allocation techniques for rate splitting based mu-mimo systems,''
\newblock {\em IEEE Transactions on Communications}, vol. 70, no. 7, pp. 4656--4670, 2022.

\bibitem{dynovs}
Zhichao Shao, Lukas T.~N. Landau, and Rodrigo~C. de~Lamare,
\newblock ``Dynamic oversampling for 1-bit adcs in large-scale multiple-antenna systems,''
\newblock {\em IEEE Transactions on Communications}, vol. 69, no. 5, pp. 3423--3435, 2021.

\bibitem{comprec}
Ana Beatriz L.~B. Fernandes, Zhichao Shao, Lukas T.~N. Landau, and Rodrigo~C. de~Lamare,
\newblock ``Multiuser-mimo systems using comparator network-aided receivers with 1-bit quantization,''
\newblock {\em IEEE Transactions on Communications}, vol. 71, no. 2, pp. 908--922, 2023.

\bibitem{zcprec}
Diana M.~V. Melo, Lukas T.~N. Landau, Rodrigo~C. de~Lamare, Peter~F. Neuhaus, and Gerhard~P. Fettweis,
\newblock ``Zero-crossing precoding techniques for channels with 1-bit temporal oversampling adcs,''
\newblock {\em IEEE Transactions on Wireless Communications}, vol. 22, no. 8, pp. 5321--5336, 2023.

\bibitem{damdc}
Tamara~Guerra Miller, Songcen Xu, Rodrigo~C. de~Lamare, and H.~Vincent Poor,
\newblock ``Distributed spectrum estimation based on alternating mixed discrete-continuous adaptation,''
\newblock {\em IEEE Signal Processing Letters}, vol. 23, no. 4, pp. 551--555, 2016.

\bibitem{rdistdiff}
Yi~Yu, Haiquan Zhao, Rodrigo~C. de~Lamare, Yuriy Zakharov, and Lu~Lu,
\newblock ``Robust distributed diffusion recursive least squares algorithms with side information for adaptive networks,''
\newblock {\em IEEE Transactions on Signal Processing}, vol. 67, no. 6, pp. 1566--1581, 2019.

\bibitem{rmultitask}
Tao Yu, Rodrigo~C. de~Lamare, and Yi~Yu,
\newblock ``Robust resilient diffusion over multi-task networks against byzantine attacks: Design, analysis and applications,''
\newblock {\em IEEE Transactions on Signal Processing}, vol. 70, pp. 2826--2841, 2022.

\end{thebibliography}

\end{document}